\documentclass[conference]{IEEEtran}
\ifCLASSINFOpdf
 \else
\fi
\usepackage{graphicx}
\usepackage{color}
\usepackage{amsmath}
\usepackage{amsfonts}
\begin{document}
\title{Sparse Message Passing Based Preamble Estimation\\for Crowded M2M Communications}
\author{\IEEEauthorblockN{Zhaoji Zhang\IEEEauthorrefmark{1}, Ying Li\IEEEauthorrefmark{1}, Lei Liu\IEEEauthorrefmark{2}, and Huimei Han\IEEEauthorrefmark{1}\\
\IEEEauthorrefmark{1}State Key Laboratory of ISN, Xidian University, Xi'an 710071, China\\
\IEEEauthorrefmark{2}City University of Hong Kong, Hong Kong, China}
E-mail:yli@mail.xidian.edu.cn
}
\maketitle
\begin{abstract}
Due to the massive number of devices in the M2M communication era, new challenges have been brought to the existing random-access (RA) mechanism, such as severe preamble collisions and resource block (RB) wastes. To address these problems, a novel sparse message passing (SMP) algorithm is proposed, based on a factor graph on which Bernoulli messages are updated. The SMP enables an accurate estimation on the activity of the devices and the identity of the preamble chosen by each active device. Aided by the estimation, the RB efficiency for the uplink data transmission can be improved, especially among the collided devices. In addition, an analytical tool is derived to analyze the iterative evolution and convergence of the SMP algorithm. Finally, numerical simulations are provided to verify the validity of our analytical results and the significant improvement of the proposed SMP on estimation error rate even when preamble collision occurs.
\end{abstract}
\IEEEpeerreviewmaketitle
\section{Introduction}
Providing efficient support for Machine-to-Machine (M2M) communications and services is one of the major objectives in the evolution of cellular networks towards the fifth generation (5G). It is anticipated that the number of M2M devices would exceed 50 billion by 2020 \cite{2020}. Attempts to access cellular networks from such a huge number of M2M devices can lead to severe congestion in the existing random-access (RA) mechanism. Furthermore, the M2M applications, such as smart city, smart metering, e-health, fleet management and  intelligent transportation, are generally characterized by small-sized data intermittently transmitted by a massive number of M2M devices. Specifically, these M2M devices would be activated with a low probability and stay off-line after data transmission \cite{servicetype}\cite{servicetypetwo}. This implies that the RA process in M2M communication exhibits prominent features of massiveness and sparseness.\\
\indent Several grant-free schemes employed compressed sensing (CS) algorithms to exploit the sparseness feature and accomplish the activity detection or the joint estimation of channel and device activity for M2M communications. For example, a block CS algorithm \cite{blockCS} was proposed for distributed device detection and resource allocation based on the clustering of devices. A greedy algorithm based on orthogonal matching pursuit was proposed in \cite{2013CS} for the user activity detection and channel estimation. The same task as in \cite{2013CS} was accomplished by a modified Bayesian compressed sensing algorithm \cite{CRAN} for the cloud radio access network. The powerful approximate message passing (AMP) algorithm \cite{CS} was studied for the joint user activity detection and channel estimation problem when the base station (BS) is equipped with one single antenna \cite{2015ICC}\cite{ICASSP} or multiple antennas \cite{ISIT}.\\
\indent As an alternative, contention-based RA schemes inherently exhibit outstanding detection performances due to the orthogonal preambles. However, the massiveness of M2M communications poses severe overload of the physical random access channel (PRACH) and preamble collisions. Several solutions were proposed for the PRACH overload \cite{ACB}-\cite{dynamicone}. Furthermore, the preamble collision was dealt with either by increasing the number of available preambles \cite{morePA2} or by reusing the preambles \cite{reusePA}. An early preamble collision detection scheme was proposed in \cite{TVTearly} based on tagged preambles which could avoid resource block (RB) wastes caused by collisions as well as monitor the RA load.\\
\indent An accurate estimation of the preambles chosen by each device offers helpful information to the uplink data transmission. Resource-friendly transmission techniques, such as NOMA, can be employed among the collided devices to improve the RB efficiency. However, to the best of our knowledge, there are few works that deal with the estimation of the collided devices. Aided by the message passing algorithm (MPA), it is possible for the base station (BS) to perform accurate estimations even if collision occurs. \\
\indent The MPA is renowned for its application in decoding low-density parity-check (LDPC) codes \cite{William2009} and CS \cite{CS}. In addition, the graph-based MPA is also applied for the Gaussian Message Passing Iterative Detector (GMPID) in massive MU-MIMO systems \cite{Lei2016Convergence} and the MIMO-NOMA systems \cite{Lei2016Gaussian}.
In this paper, we propose a sparse message passing (SMP) algorithm targeting on estimating the user-preamble indicator matrix at the first step of the RA process. Although it is similar to the Belief Propagation (BP) decoder \cite{William2009} and GMPID \cite{Lei2016Convergence}, the SMP exhibits the following major differences and advantages.\\
\indent The SMP inherits the low complexity of the MPA by departing the overall processing into distributed calculations that can be executed in parallel. Bernoulli messages are updated on the factor graph, which is different from the GMPID. Furthermore, check nodes (CNs) and sum nodes (SNs) of the SMP follow different update rules from BP decoders. Despite the fact that the RA estimation is a special case of the sparse recovery problem, the major difference between the SMP and other existing solutions such as the AMP is that the SMP algorithm considers the CN constraint in every iteration, i.e., only one preamble can be chosen by an activated device. Furthermore, the AMP algorithm may not be preferred for large dynamic systems since an inappropriately tuned threshold function may lead to sharply deteriorated performance \cite{threshold}.\\
\indent The contributions of this paper are summarized as follows.\\
\indent (i) A SMP algorithm is proposed for the RA system to estimate the activity of the devices and the identity of the preamble chosen by each active device.\\
\indent (ii) A factor graph for the SMP is presented to illustrate the message update rules for different type of nodes.\\
\indent (iii) An analytical tool is derived to analyze the iterative evolution of the messages, which illustrates the impacts of system parameters and the convergence of the SMP algorithm.\vspace{-0.2cm}\\
\section{System Model}
We consider a single cell centered by a BS in a cellular wireless network with the following assumptions. The number of M2M devices is far greater than that of human-to-human (H2H) devices in the cell.  The BS is equipped with $M$ antennas while each device is with one single antenna. The channel between each device and the BS is a slow time-varying block fading TDD (time division duplex) channel, so that the uplink channel matrix is identical to that of the downlink. There are $N_s$  devices, each of which is activated with probability $p_a$. $N_p$ preambles with length $N_c$ are assigned to the RA system. At the start of a RA slot, each active device randomly selects a preamble with equal probability $1/N_p$.\\
\indent The received signal of the $l$-th antenna at the time instant $t$ can be written as
\begin{equation}\label{model}
y_l^t = \sum\limits_{i = 0}^{{N_s} - 1} {\sum\limits_{j = 0}^{{N_p} - 1} {{h_{li}}{s_{ij}}P_j^t} }  + n_l^t,
\end{equation}
where $h_{li}$ is the channel gain from the $i$-th device to the $l$-th antenna,  $P_j^t$ is the $t$-th symbol of the $j$-th preamble for $t=0,1,\ldots, N_c-1$ and $s_{ij}$ is the user-preamble indicator, i.e., $s_{ij}=1$ if the $i$-th device selects the $j$-th preamble. Otherwise, $s_{ij}=0$. Rewriting (\ref{model}) in a matrix form, we have
\begin{equation}\label{modelmatrix}
{\bf{Y}}_{M\times N_c} = { \bf{H} }_{M\times N_s}{ \bf{S} }_{N_s\times N_p}{ \bf{P} }_{N_p\times N_c} + {\bf{N}}_{M\times N_c},
\end{equation}
where $\mathbf{N}$ is an \emph{i.i.d} additive Gaussian noise matrix with variance $\sigma_n^2$. $\mathbf{S}$ is an user-preamble indicator matrix. The probability that a certain row of $\mathbf{S}$ has just one \lq\lq1\rq\rq\ is $p_a$ while the probability for each entry $s_{sp}$ to be 1 is
\begin{equation}
{p_0}=Pr\left(s_{sp}=1\right)=p_a/N_p.
\end{equation}
$\mathbf{P}$ is an $N_p\times N_c$ preamble matrix, each row of which represents a preamble. Generally, Zadoff-Chu sequences \cite{TVTearly} are employed as RA preambles due to their ideal \emph{auto-correlation property}, i.e., ${\bf{P}}{{\bf{P}}^H} = {N_c}{\bf{I}}$, where $\mathbf{I}_{{N_P} \times {N_P}}$ is an identity matrix. With this property, we can rewrite (\ref{modelmatrix}) to
\begin{equation}
\bar{\mathbf{Y}}\! =\! N_c^{-1}\mathbf{Y}\mathbf{P}^H \! =\! N_c^{-1}\mathbf{H}\mathbf{S}\mathbf{P}\mathbf{P}^H \!+\! N_c^{-1}\mathbf{N}\mathbf{P}^H \! =\! \mathbf{H}\mathbf{S} \!+\! \bar{\mathbf{N}}.
\end{equation}
\indent The system model can be further simplified by introducing a little ambiguity,
\begin{equation}\label{new_sys_model}
{\mathbf{Y}}_{M\times N_p} = \mathbf{H}_{M\times N_s}\mathbf{S}_{N_s\times N_p} + {\mathbf{N}}_{M\times N_p},
\end{equation}
where ${\mathbf{N}}\sim\mathcal{N}^{M\times N_p}(0,\sigma_n^2/N_c)$. The channel matrix $\mathbf{H}$ is assumed known to the BS, which is a practical assumption in the TDD-based wireless access system. Therefore $\mathbf{S}$ is the only target for our estimation.\\
\section{Message Passing for Indicator Matrix Estimation}
\begin{figure}
  \centering
  \includegraphics[width=6.5cm]{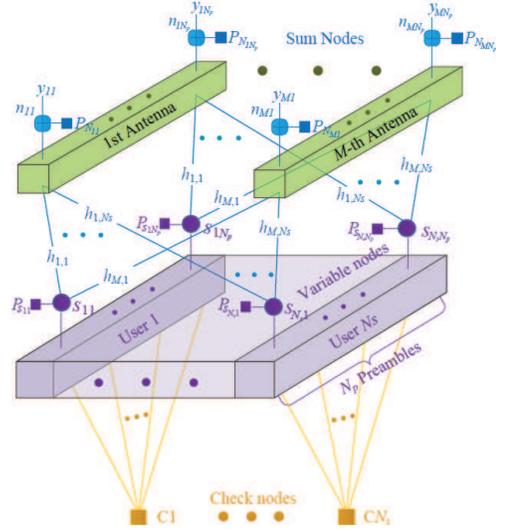}\vspace{-0.2cm}\\
  \caption{Factor graph of sparse message passing for random-access estimation.}\label{f1}\vspace{-0.3cm}
\end{figure}
In Fig. \ref{f1}, we present a factor graph for the SMP algorithm, on which Bernoulli messages are updated for each entry in the indicator matrix $\mathbf{S}$. As shown, a SN $y_{mp}$ is the received signal of the $m$-th antenna with respect to the $p$-th preamble, a variable node (VN) $s_{sp}$ represents the $p$-th entry of the $s$-th row in $\mathbf{S}$ while the CNs stand for the check node constraints for each device. The message updating diagram among SNs, CNs and VNs is illustrated in Fig. \ref{f2}. The output message is defined as the extrinsic information which, according to the rule of message passing, is derived by the incoming messages from the other edges that are connected to the same node.
\subsection{Message Update at Sum Nodes}
\begin{figure*}
  \centering
  \includegraphics[width=1\textwidth]{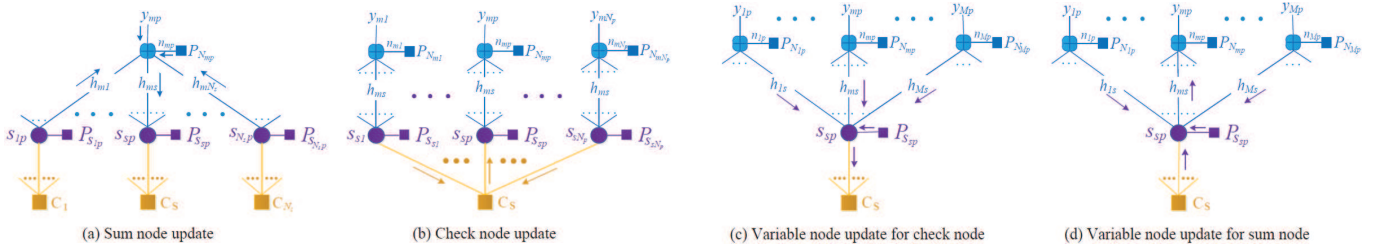}\vspace{-0.3cm}\\
  \caption{Message update at the sum nodes, check nodes and variable nodes. The message passed on each edge are the non-zeros probabilities of the Bernoulli variables in the user-preamble indicator matrix $\mathbf{S}$. The output message called extrinsic information is derived by the messages on the other edges that are connected with the same node.}\label{f2}\vspace{-0.3cm}
\end{figure*}

Each SN can be seen as a multiple-access process and the message update at the $mp$-th SN for the $sp$-th VN is presented as an example in Fig. \ref{f2}(a). Firstly, the received signal $y_{mp}$ at the $mp$-th SN can be rewritten to
\begin{eqnarray}
y_{mp}&=&{h_{ms}s_{sp}} + \sum\limits_{i \in {{\cal N}_s}/s} {{h_{mi}}{s_{ip}}}  + {n_{mp}}\nonumber\\
&=&{h_{ms}s_{sp}} +n^*_{mps}(\tau).\label{expandy}
\end{eqnarray}
where $s\in \mathcal{N}_s$ for $\mathcal{N}_s =\{1,\ldots,N_s\}$, $m\in\mathcal{M}$ for $\mathcal{M}=\{1,\ldots, M\}$ and $p\in\mathcal{N}_p$ for $\mathcal{N}_p=\{1,\ldots,N_p\}$. We assume that $p_{sp\to mp}^{vs}(\tau)$ denotes the non-zero probability for the Bernoulli variable $s_{sp}$ passing from the $sp$-th VN to the $mp$-th SN in the $\tau$-th iteration. Based on the \emph{central limit theorem}, $n_{mps}^*(\tau)$ can be approximated as an equivalent Gaussian noise with mean $u_{mps}^*(\tau)$ and variance $v_{mps}^*(\tau)$,
\begin{equation}\label{equ_noi}
\left\{ \begin{array}{l}
\!u_{mps}^*(\tau)\! =\! \sum\limits_{i\in {\cal N}_s/s} {\!\!\!h_{mi}}p_{ip \to mp}^{vs}(\tau),\\
\!v_{mps}^*(\tau)\! =\! \sum\limits_{i\in {\cal N}_s/s} {\!\!\!h_{mi}^2p_{ip \to mp}^{vs}(\tau)q_{ip\! \to\! mp}^{vs}(\tau)\!+\!\sigma _n^2},
\end{array} \right.
\end{equation}
where $p_{ip\to mp}^{vs}(\tau)$ and $q_{ip \to mp}^{vs}(\tau)$ are the non-zero and zero probabilities for the Bernoulli variable $s_{ip}$ passed from the $ip$-th VN to the $mp$-th SN. Then the extrinsic message $p_{mp\to sp}^s(\tau)$ from the $mp$-th SN to the $sp$-th VN is
\begin{equation}\label{sum_Ber}
\begin{array}{l}
\displaystyle p_{mp \to sp}^s(\tau ) = {\left[ {1 + \frac{{Pr\left( {{s_{sp}} = 0|{y_{mp}},{\textbf{h}_m},p_{mp,\sim sp}^{vs}(\tau )} \right)}}{{Pr\left( {{s_{sp}} = 1|{y_{mp}},{\textbf{h}_m},p_{mp,\sim sp}^{vs}(\tau )} \right)}}} \right]^{ - 1}}\\
\displaystyle ={\left[ {1 + \frac{{Pr\left( {{y_{mp}} = n_{mps}^*(\tau )|u_{mps}^*(\tau ),v_{mps}^*(\tau )} \right)}}{{Pr\left( {{y_{mp}} = {h_{ms}} + n_{mps}^*(\tau )|u_{mps}^*(\tau ),v_{mps}^*(\tau )} \right)}}} \right]^{ - 1}}\\
\displaystyle = {\left[ {1 + \frac{{f\left( {{y_{mp}}|u_{mps}^*(\tau ),v_{mps}^*(\tau )} \right)}}{{f\left( {{y_{mp}}|u_{mps}^*(\tau ) + {h_{ms}},v_{mps}^*(\tau )} \right)}}} \right]^{ - 1}}.
\end{array}
\end{equation}
where $\mathbf{h}_m$ is the $m$-th row of $\mathbf{H}$, ${\bf{p}}_{mp,\sim sp}^{vs}(\tau)$ is the set of Bernoulli probabilities $\{{{p}}_{ip\to mp}^{vs}(\tau)|i\in \mathcal{N}_s/s\}$ and $f(x|u,v)$ is the \emph{probability density function} of a Gaussian distribution $\mathcal{N}(u,v)$, i.e.,
\begin{equation}\label{Gaussianpdf}
  f\left(x|u,v \right)=\frac{1}{\sqrt {2\pi v}}e^{-\frac{(x-u)^2}{2v}}.
\end{equation}
\indent To avoid overflow caused by a large number of multiplications of probabilities, we put the Bernoulli messages in the form of \emph{log-likelihood ratio} (LLR).
\begin{equation}\label{SUMNODE}
\displaystyle l_{mp \to sp}^s (\!\tau\!)\!\! = \!\!\log \frac{{p_{mp \to sp}^s(\!\tau\!)}}{{q_{mp \to sp}^s(\!\tau\!)}}\!\!=\!\!{\frac{2(y_{mp}\!\!-\!\!u_{mps}^* (\!\tau\!))h_{ms}\!\!-\!\!h_{ms}^2}{2v_{mps}^*(\!\tau\!)\!}}\!.
\end{equation}
\subsection{Message Update at Variable Nodes}
Each VN can be seen as a broadcast process and the extrinsic message from the VN is derived following the message combination rule \cite{Loeliger2006}, i.e., the extrinsic message is a normalized product of the input probabilities.
\subsubsection{Message update for the sum nodes}
The message update from the $sp$-th VN to the $mp$-th SN is presented as an example in Fig. \ref{f2}(d). The extrinsic message is derived by the initial probability ${\bar p}_{sp}$ of each VN, the incoming message from the $s$-th CN $p_{s\to sp}^c(\tau )$ and the messages from the $jp$-th SN with $j\in\mathcal{M}/m$. According to the message combination rule, the extrinsic message from the $sp$-th VN to the $mp$-th SN is
\begin{eqnarray}\label{var_ber1}
\begin{array}{l}
\displaystyle p_{sp \to mp}^{vs}(\tau \!\! + \!\!1)\!\! = \!\!Pr({s_{sp}=1}|\textbf{p}_{sp,\sim mp}^s(\tau ),p_{s \to sp}^c(\tau ),{{\bar p}_{sp}})\vspace{0.2cm}\\
\displaystyle \!\!\mathop  = \limits^{(a)} \!\!\frac{{{{\bar p}_{sp}}p_{s \to sp}^c(\tau )\!\!\!\prod\limits_{j \in {\cal M}/m} \! {p_{jp \to sp}^s}(\tau ) }}{{{{\bar p}_{sp}}p_{s \to sp}^c\!(\!\tau\! )\!\!\!\!\!\prod\limits_{j \in {\cal M}/m}\!\!\!\! {p_{jp \to sp}^s}\!(\!\tau\! )  \!+\! {{\bar q}_{sp}}q_{s \to sp}^c\!(\!\tau\! )\!\!\!\!\!\prod\limits_{j \in {\cal M}/m}\!\! \!{q_{jp \to sp}^s}\!(\!\tau \!) }}\vspace{0.2cm}\\
\displaystyle \!\!\mathop  = \limits^{(b)} \!\!\frac{{{{p}_{0}}p_{s \to sp}^c(\tau )\!\!\!\prod\limits_{j \in {\cal M}/m} \! {p_{jp \to sp}^s}(\tau ) }}{{{{p}_{0}}p_{s \to sp}^c(\!\tau\! )\!\!\!\!\!\prod\limits_{j \in {\cal M}/m}\!\!\!\! {p_{jp \to sp}^s}(\!\tau\! )  \!+\! {{q}_{0}}q_{s \to sp}^c(\!\tau\! )\!\!\!\!\!\prod\limits_{j \in {\cal M}/m}\!\! \!{q_{jp \to sp}^s}(\!\tau \!) }}\vspace{-0.05cm},\\
\end{array}
\end{eqnarray}
where ${\bf{p}}_{sp,\sim mp}^s(\tau)=\{p^s_{jp\to sp}(\tau)|j\in\mathcal{M}/m\}$. Equation ($a$) is derived by the normalized product of the input Bernoulli probabilities and ($b$) is obtained by $\bar{p}_{sp}= p_0$ and $\bar{q}_{sp}= q_0$.
\subsubsection{Message update for the check nodes}
Similarly, in Fig. \ref{f2}(c), the extrinsic message update of $s_{sp}$ from the $sp$-th VN to the $s$-th CN is derived by the initial probability ${\bar p}_{sp}$ and the messages passing from the $mp$-th SN where $m\in\mathcal{M}$.
\begin{eqnarray}\label{var_ber2}
\begin{array}{l}
\displaystyle p_{sp \to s}^{vc}(\tau  + 1) = Pr({s_{sp}=1}|\textbf{p}_{sp}^s(\tau ),{{\bar p}_{sp}})\vspace{0.2cm}\\
\displaystyle \mathop  = \limits^{(c)} \frac{{{{\bar p}_{sp}}\prod\limits_{m \in {\cal M}} {p_{mp \to sp}^s(\tau )} }}{{{{\bar p}_{sp}}\prod\limits_{m \in {\cal M}} {p_{mp \to sp}^s} (\tau ) + {{\bar q}_{sp}}\prod\limits_{m \in {\cal M}} {q_{mp \to sp}^s(\tau )} }}\vspace{0.2cm}\\
\displaystyle \mathop  = \limits^{(d)} \frac{{{{p}_0}\prod\limits_{m \in {\cal M}} {p_{mp \to sp}^s(\tau )} }}{{{{p}_0}\prod\limits_{m \in {\cal M}} {p_{mp \to sp}^s(\tau )}  + {{q}_0}\prod\limits_{m \in {\cal M}} {q_{mp \to sp}^s(\tau )} }},\vspace{-0.05cm}
\end{array}
\end{eqnarray}
where $\textbf{p}_{sp}^s(\tau)=\{p_{mp\to sp}^s(\tau)|m\in\mathcal{M}\}$. Equation ($c$) is derived by the normalized product of the input Bernoulli probabilities and ($d$) is obtained by $\bar{p}_{sp}= p_0$ and $\bar{q}_{sp}= q_0$.\\
\indent Similarly, the extrinsic LLR messages from the VNs are
\begin{eqnarray}\label{SDvnupdate}
\left\{ \begin{array}{l}
l_{sp \to mp}^{vs}(\tau  + 1) =\! \!\!\sum\limits_{j \in \mathcal{M}/m}\!\! {l_{jp \to sp}^s(\tau )}  + {{\bar l}_{sp}} + l_{s \to sp}^c(\tau ),\vspace{0.2cm}\\
l_{sp \to s}^{vc}(\tau  + 1) = \sum\limits_{m \in \mathcal{M}} {l_{mp \to sp}^s(\tau )}  + {{\bar l}_{sp}},
\end{array} \right.
\end{eqnarray}
with $l_{s\to sp}^c(\tau)\!=\!\log(p_{s\to sp}^c(\tau)/q_{s\to sp}^c(\tau))$, ${\bar l}_{sp}\!=\!\log({\bar p}_{sp}/{\bar q}_{sp})$ and $l_{jp\to sp}^s(\tau)\!=\!\log(p_{jp\to sp}^s(\tau)/q_{jp\to sp}^s(\tau))$.
\subsection{Message Update at Check Nodes}
The $s$-th CN represents a constraint for the corresponding VNs that an VN $s_{sp}=1$ if and only if the $s$-th device is active and  the other VNs $s_{sk}=0$ for any $k\in \mathcal{N}_p/p$. As illustrated in Fig. \ref{f2}(b), the extrinsic message from the $s$-th CN to the $sp$-th VN is derived by the initial activation probability $p_a$ for this device and the incoming messages from the $sk$-th VN with $k\in \mathcal{N}_p/p$. The message update is presented as
\begin{eqnarray}\label{check_update}
\displaystyle p_{s \to sp}^{c}(\tau)\!=\! Pr(s_{sp}\!=\!1|{p_a,}\; \mathbf{p}_{s,\sim p}^{vc}(\!\tau\!))\!=\!{p_a}\!\!\mathop \Pi \limits_{k\in \mathcal{N}_p\!/\!p}\!\! q_{sk\to s}^{vc}(\!\tau\!),
\end{eqnarray}
where $\mathbf{p}_{s,\sim p}^{vc}(\tau)=\{{{p}}_{sk\to s}^{vc}(\tau)|k\in \mathcal{N}_p/p\}$. The extrinsic LLR message $l_{s\to sp}^c(\tau)$ from the $s$-th CN is derived by
\begin{equation}\label{CN2VN}
\begin{array}{l}
\displaystyle \tilde l_{s \to sp}^c(\tau ) = \log\left( {p_{s \to sp}^c(\tau )} \right)\vspace{0.2cm}\\
\displaystyle  = \log ({p_a}) - \sum\limits_{k \in {{\cal N}_p}/p} {\log (} {e^{l_{sk \to s}^{vc}(\tau )}} + 1),
\end{array}
\end{equation}
\begin{equation}\label{check_update_LLR}
\!\!\!\!\!\!\!\!l_{s \to sp}^{c}(\tau)= -\log\big( e^{-\tilde{l}_{s \to sp}^{c}(\tau)} - 1\big).
\end{equation}
\subsection{Output and Decision}
The final output Bernoulli message for each VN is derived by all the incoming messages.
\begin{eqnarray}\label{var_dec}
\hat{p}_{sp}\!(\!\tau\!)\!\!=\!\! \frac{{{p_0}\!\,p_{s\to sp}^c\!(\!\tau\!)\!\!\!\!\mathop \Pi \limits_{m \in \mathcal{M}} \!\!\!p_{mp \to sp}^s\!(\!\tau\!)}}{{{p_0}\!\,p_{s\to sp}^c\!(\!\tau\!)\!\!\!\!\mathop \Pi \limits_{m \in \mathcal{M}} \!\!\!p_{mp \to sp}^s\!(\!\tau\!) \!\!+\!\! q_0\!\,q_{s\to sp}^c\!(\!\tau\!)\!\!\!\!\mathop \Pi \limits_{m \in \mathcal{M}} \!\!\!q_{mp \to sp}^s\!(\!\tau\!)}}\!.
\end{eqnarray}
The output LLR message for each Bernoulli variable $s_{sp}$ is
\begin{eqnarray}\label{var_dec_LLR}
\hat{l}_{sp}(\tau) = \mathop \sum \limits_{m \in \mathcal{M}} l_{mp \to sp}^s(\tau)+ \bar{l}_{sp} + l_{s\to sp}^c(\tau).
\end{eqnarray}
Then the final decision for each Bernoulli variable is given by
\begin{equation}\label{zerone}
\hat{s}_{sp} =\left\{ \begin{array}{l}
1, \;\;\mathrm{if}\;\; \hat{l}_{sp}\geq0 \\
0, \;\;\mathrm{if}\;\; \hat{l}_{sp}<0
\end{array} \right.
\end{equation}
Therefore, $\widehat{\mathbf{S}}=[\hat{{s}}_{sp}]_{N_s\times N_p}$ is the final estimate.
\section{Analysis for Iterative Evolution}
In this section, the iterative evolution of the messages passed among the SNs, VNs and CNs are analyzed while each node in the factor graph can be considered as a processor. The analysis is analogous to the density evolution (DE) \cite{William2009} which is widely employed for analyzing LDPC codes in the asymptotic regime. Furthermore, we take the blunt assumption that the LLR messages are independently and identically distributed (\emph{i.i.d}). Unfortunately, the Gaussian assumption (GA) and consistency condition employed to simplify the DE analysis do not hold for our system model. As an alternative, only the mean of the LLR messages is tracked, which could still guarantee an effective insight for the performance of the proposed SMP algorithm. Unlike the DE for LDPC codes where all transmitted bits are assumed to be +1, there are two types of VNs, i.e. the chosen VNs ($s_{sp}=1$) and those that are not chosen ($s_{sp}=0$). The chosen VNs are referred to as \lq\lq positive\rq\rq\ since the outcoming and incoming LLR messages for such VN processors are assumed positive while the other VNs are referred to as \lq\lq negative\rq\rq\ for similar reasons.
\begin{figure*}
  \centering
  \includegraphics[width=0.83\textwidth]{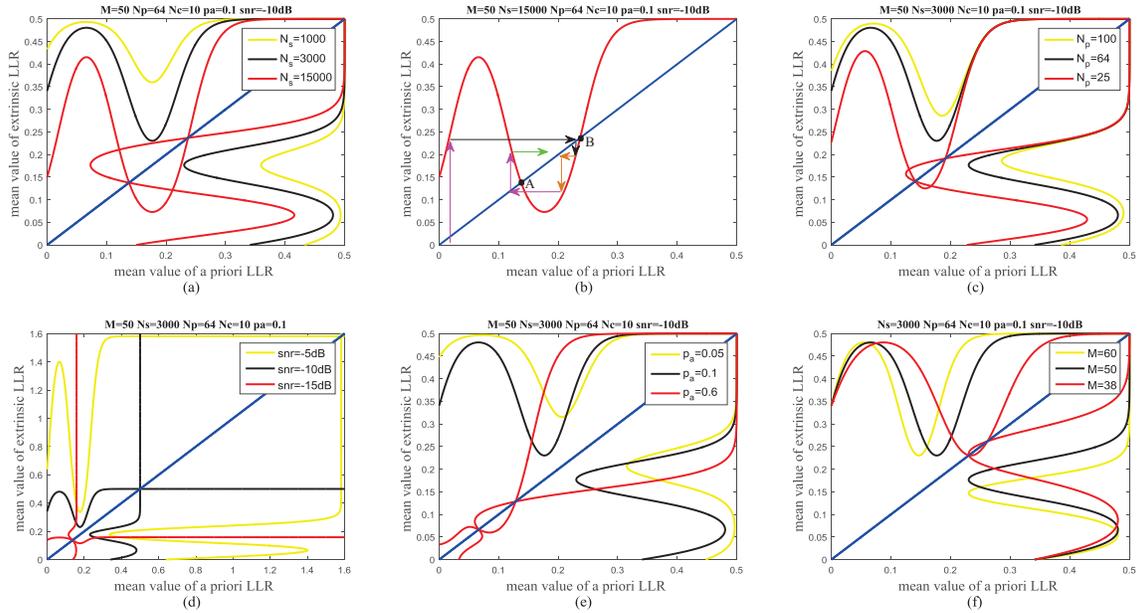}\vspace{-0.3cm}\\
  \caption{Numerical results on the impacts of system parameters $N_s$, $N_p$, SNR, $p_a$ and $M$.}\vspace{-0.2cm}\label{numericalll}\vspace{-0.3cm}
\end{figure*}
\subsection{Analysis for Variable Node Processors}
According to the \emph{i.i.d} assumption, we derive the expectation of equation (\ref{SDvnupdate}) while omitting the original subscript. For positive and negative VN processors, we have
\begin{equation}\label{VNPp2SNPCNP}
\left\{ \begin{array}{l}
\displaystyle E[l_ + ^{vs}(\tau)] = (M - 1)E[l_ + ^s(\tau )] + {{\bar l}_{sp}} + E[l_ + ^c(\tau )],\vspace{0.05cm}\\
\displaystyle E[l_ + ^{vc}(\tau)] = ME[l_ + ^s(\tau )] + {{\bar l}_{sp}}.
\end{array} \right.
\end{equation}
\begin{equation}\label{VNPn2SNPCNP}
\left\{ \begin{array}{l}
\displaystyle E[l_ - ^{vs}(\tau )] = (M - 1)E[l_ - ^s(\tau )] + {{\bar l}_{sp}} + E[l_ - ^c(\tau )],\vspace{0.05cm}\\
\displaystyle E[l_ - ^{vc}(\tau )] = ME[l_ - ^s(\tau )] + {{\bar l}_{sp}}.
\end{array} \right.
\end{equation}
\indent Note that the subscripts $+$ and $-$ indicate the VN type. $E[l^{vs}(\tau)]$ and $E[l^{vc}(\tau)]$ are the mean of the LLR messages passing from VNs to SNs and CNs while $E[l^s(\tau)]$ and $E[l^c(\tau)]$ are the mean of the LLR messages passed to VNs from SNs and CNs respectively.
\subsection{Analysis for Check Node Processors}
The message passing from an active CN processor to a positive VN is derived by the incoming messages from $N_p-1$ negative VNs. Therefore, according to (\ref{CN2VN}) and (\ref{check_update_LLR}), we have
\begin{equation}\label{CN2VNPp}
\displaystyle E[\tilde l_ + ^c(\tau )] \approx \log ({p_a}) - ({N_p} - 1)\log ({e^{E[l_ - ^{vc}(\tau )]}} + 1),
\end{equation}
\begin{equation}\label{CN2VNPall}
\displaystyle E[l_ + ^c(\tau )] \approx  - \log({e^{ - E[\tilde l_ + ^c(\tau )]}} - 1).
\end{equation}
\indent The message passing from an active CN processor to a negative VN is derived by the incoming messages from $N_p-2$ negative VNs and 1 positive VN,\\
\begin{equation}\label{CN2VNnac}
\left\{ \begin{array}{l}
\displaystyle\!\!\!\! E[\tilde l_{ \!- ac}^c(\!\tau\! )\!]\!\approx\!\log ({p_a}\! ) \!\!- \!\!(\! {N_p} \!\!- \!\!2\! )\!\log(\! {e^{E[l_ - ^{vc}\! (\tau )]}} \!\!+ \!\!1\! ) \!\!- \!\!\log (\! {e^{E[l_ + ^{vc}\! (\tau )]}} \!\!+ \!\!1\! ),\vspace{0.05cm}\\
\displaystyle\!\!\!\! E[l_{- ac}^c\!(\!\tau\! )\!] \!\approx\! - \log(\! {e^{ - E[\tilde l_{ - ac}^c(\tau )]}} - 1),
\end{array} \right.
\end{equation}where \lq\lq$-ac$\rq\rq\ indicates that the CN is active while the VN is negative. The LLR passed to a negative VN from an inactive CN is derived by the messages from $N_p-1$ negative VNs,
\begin{equation}\label{CN2VNnina}
\left\{ \begin{array}{l}
\displaystyle \!\!\!\!E[\tilde l_{ - ina}^c(\tau )] \!\approx\!\log ({p_a}) \!- \!({N_p} \!- \!1)\log ({e^{E[l_ - ^{vc}(\tau )]}} \!+\! 1),\vspace{0.1cm}\\
\displaystyle \!\!\!\!E[l_{ - ina}^c(\tau )] \!\approx\!  - \log({e^{ - E[\tilde l_ {-ina} ^c(\tau )]}} \!- \!1),
\end{array}\right.
\end{equation}where \lq\lq$-ina$\rq\rq\ indicates that the CN is inactive while the VN is negative. The mean of the LLR messages from CNs to negative VNs is averaged by the activation probability $p_a$,
\begin{equation}\label{CN2VNn_ave}
E[l_{ -}^c(\tau )] = {p_a}E[l_{ - ac}^c(\tau )] + (1 - {p_a})E[l_{ - ina}^c(\tau )].
\end{equation}
\subsection{Analysis for Sum Node Processors}
We assume that the channel gain is \emph{i.i.d}, i.e., ${h_{mi}}\sim{\cal N}(0,1)$ and derive the expectation of (\ref{SUMNODE}) for a positive VN.
\begin{equation}\label{firstsum}
\begin{array}{l}
\displaystyle\!\! E[l_ + ^s(\tau  + 1)]\vspace{0.05cm}\\
\displaystyle\!\! \mathop  = \limits^{(e)} E\left[ {\frac{{2({h_{ms}}{s_{sp}} - \sum\limits_{i \in {{\cal N}_s}/s} {{h_{mi}}p_{ip \to mp}^{vs}(\tau )} ){h_{ms}} - h_{ms}^2}}{{2\sum\limits_{i \in {{\cal N}_s}/s} {h_{mi}^2p_{ip \to mp}^{vs}(\tau )(1 - p_{ip \to mp}^{vs}(\tau )) + 2\sigma _n^2} }}} \right]\vspace{0.05cm}\\
\displaystyle\!\! \mathop  = \limits^{(f)} E\left[ {{{\left( {\sum\limits_{i \in {{\cal N}_s}/s}\!\!{\frac{2}{{2 \!+\! {e^{ - l_{ip \to mp}^{vs}(\tau )}} \!+ \!{e^{l_{ip \to mp}^{vs}(\tau )}}}} \!\!+ \!\!2\sigma _n^2} } \right)}^{ - 1}}} \right]\vspace{0.05cm}\\
\displaystyle\!\!  \mathop  \approx \limits^{(g)}\!\! {\left( {\frac{{2({N_s} - 1){p_0}}}{{2 \!+\! {e^{\! -\! E[l_ + ^{v\!s}\!(\tau )]\!}}\! + \!{e^{\!E[l_ + ^{v\!s}\!(\tau )]}}}}\!\! +\!\! \frac{{2({N_s} - 1)(1 - {p_0})}}{{2 \!+ \!{e^{ \!- \!E[l_ - ^{v\!s}\!(\tau )]}} \!+ \! {e^{\!E[l_ - ^{v\!s}\!(\tau )]}}}}\! \!+\! \!2\sigma _n^2} \!\right)^{ \!\!- 1\!\!\!}}\!\!.
\end{array}
\end{equation}
where $(e)$ is obtained from (\ref{equ_noi}) and the distributions of $h_{mi}$ and $n_{mp}$, $(f)$ is obtained from the facts that $s_{sp}\!=\!1$ for positive VNs, $E[h_{mi}^2] = 1$ and $p_{sp \to mp}^{vs}(\!\tau\!)=1/(1+e^{-l_{sp \to mp}^{vs}(\!\tau\!)})$. Approximation ($g$) is derived from the fact that the $N_s-1$ independent incoming messages $l_{ip \to mp}^{vs}(\tau)$ are passed from $N_s-1$ VNs, each of which is positive with probability $p_0$ and negative with probability $1-p_0$.\\
\indent We can derive $E[l_ - ^s(\tau  + 1)]$ for negative VNs by substituting $s_{sp}=0$ into (\ref{firstsum}) and observe the symmetry property,
\begin{equation}\label{opposite}
\displaystyle E[l_ - ^s(\tau  + 1)]=-E[l_ + ^s(\tau  + 1)].
\end{equation}
\indent Finally a generalization of (\ref{VNPp2SNPCNP}), (\ref{VNPn2SNPCNP}), (\ref{CN2VNPall}), (\ref{CN2VNn_ave}), (\ref{firstsum}) and (\ref{opposite}) would complete our analysis on the iterative evolution.
\subsection{Impacts of System Parameters}
According to (\ref{opposite}), $l_{mp \to sp}^s$ is chosen to demonstrate the message evolution due to its symmetry for positive and negative VNs. The numerical results in Fig. \ref{numericalll} are illustrated in the form of an extrinsic information transfer (EXIT) chart \cite{William2009} where the a priori and extrinsic LLRs refer to the input and output of a SN processor respectively.
\subsubsection{Impacts of $N_s$, $N_p$ and $p_a$}
$N_s$, $N_p$ and $p_a$ influence the device activity or collision probability, and thus they pose similar impacts on the SMP performance. As depicted in Fig. \ref{numericalll}(a), (c) and (e) respectively, when the system is not congested, an open tunnel exists for the extrinsic LLR to converge to the upper-right point. However, severe congestion may close the tunnel and cause divergence. To elaborate on the divergent case, the extrinsic LLR curve for $N_s=15000$ in Fig. \ref{numericalll}(a) is replotted in Fig. \ref{numericalll}(b) where arrows in different colors indicate an endless loop of the successive iterations. Therefore we conclude that divergence occurs if i) an interval [A,B] exists where the extrinsic LLR curve lies below the line $y=x$ and ii) the maximal value before point A is greater than A.
\subsubsection{Impacts of SNR and $N_c$}
According to (\ref{new_sys_model}), the equivalent Gaussian noise matrix is ${\mathbf{N}}\sim \mathcal{N}^{M\times N_p}(0,\sigma_n^2/N_c)$. Therefore SNR (i.e. $\sigma_n^2$) and $N_c$ impact the estimation performance in a similar way, as shown in Fig. \ref{numericalll}(d). Note that, unlike $N_s$, $N_p$ and $p_a$, SNR affects the value of the fixed point, which can be explained via the approximation ($g$) in (\ref{firstsum}) since $2\sigma^2$ would be the dominant term after several iterations.
\subsubsection{Impact of $M$}
A larger number of antennas could offer more information to the VN processor. The impact of $M$ is depicted in Fig. \ref{numericalll}(f), from which we observe that the extrinsic LLR could converge to the same fixed point as long as the tunnel is open. Furthermore, the output LLR at the VN processor grows almost linearly with $M$ according to (\ref{VNPp2SNPCNP}), which could improve the performance.
\section{Simulation Results}
\begin{figure*}
  \centering
  \includegraphics[width=0.83\textwidth]{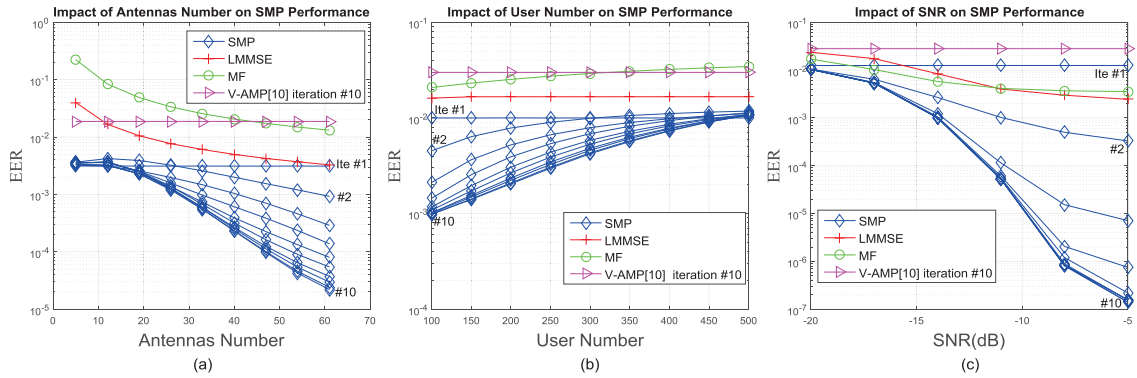}\vspace{-0.35cm}\\
  \caption{Simulation results for the performance of the proposed SMP algorithm. Related parameters are (a) $M$=5:7:61, $N_s$=300, $N_p$=64, $N_c$=10, SNR=-10dB and $p_a=0.2$ (b) $M$=30, $N_s$=100:50:500, $N_p$=50, $N_c$=10, SNR=-10dB and $p_a=0.5$ (c) $M$=60, $N_s$=100, $N_p$=64, $N_c$=10, SNR=--20:3:-5dB and $p_a=0.8$. The entries in the channel matrix $\mathbf{H}$ are assumed \emph{i.i.d} with distribution ${h_{mi}}\sim{\cal N}(0,1)$.}\vspace{-0.45cm}\label{simulation}
\end{figure*}
In this section, we simulate the performance of the SMP algorithm. Note that by EER we mean the estimation error rate for every entry of $\mathbf{S}$. The impacts of the system parameters $M$, $N_s$ and SNR are investigated in Fig. \ref{simulation}. For comparison, the performances of the linear minimum mean square error (LMMSE) estimator, the matched filter (MF) estimator and the vector AMP (V-AMP) estimator \cite{ISIT} are also included with additional check node constraint.\\
\indent The impact of $M$ is shown in Fig. \ref{simulation}(a). It can be seen that when $M$ is small, e.g., $M$=5 and 12, the SMP algorithm is inevitably divergent. However, when $M$ is a little larger, e.g. $M=19$ and 26, the performance will get improved after the second iteration, indicating that the divergent interval [A,B] is small enough for the extrinsic LLR to jump out of after several iterations. Furthermore, the divergent interval will vanish if $M$ is sufficiently large and then convergence can be guaranteed.\\
\indent Similar observations can be found in Fig. \ref{simulation}(b) for the impact of $N_s$. It is shown that the convergence can be accelerated when the value of $N_s$ becomes smaller since the evolution tunnel is more widely opened.\\
\indent The impact of SNR is shown in Fig. \ref{simulation}(c). Obviously, the increase of SNR shows the most prominent improvement for the EER performance since, as the iteration proceeds, $2\sigma^2$ becomes the dominant term in ($g$) of (\ref{firstsum}). Again this observation is consistent with the analytical results in Fig. \ref{numericalll}(d).\\
\indent It can be observed from Fig. \ref{simulation} that, under different settings, the SMP algorithm outperforms the MF estimator and the LMMSE estimator which requires matrix inversion. Since $\mathbf{S}$ follows a different distribution from the estimation target in \cite{ISIT}, the inappropriate threshold function leads to the detection failure of the V-AMP estimator. Despite the fact that simulation parameters in Fig. \ref{simulation} indicate high preamble collision probabilities,  the SMP algorithm demonstrates remarkable estimation accuracy within a feasible number of iterations.  Therefore, the RB wastes caused by collided devices can be potentially avoided with the accurate estimation of the SMP.
\section{Conclusion}
A SMP algorithm based on message passing was proposed for the RA estimation in M2M scenarios. We presented a factor graph representation for the SMP algorithm on which Bernoulli messages are updated. The message update rules were elaborated for different types of nodes. The analytical tool derived in this paper enables a visual expression for the message evolution and the convergence of the SMP algorithm. Finally simulation results verified the validity of the analysis and the outstanding accuracy of the SMP algorithm even when preamble collision occurs.

\end{document}